\documentclass[twocolumn,secnumarabic,amssymb, nobibnotes, aps, prc,superscriptaddress, showpacs, showkeys, floatfix,nofootinbib]{revtex4-1}
\usepackage{graphicx}
\usepackage{array,gensymb}
\usepackage{csvsimple,longtable,booktabs}
\usepackage{multirow}
\usepackage{adjustbox,verbatim}
\usepackage{hyperref,isotope,amsmath}
\usepackage{comment}

\begin{document}

\preprint{APS/123-QED}
\title{\boldmath Proton-unbound states in \isotope[24]Al relevant for the \isotope[23]Mg$(p,\gamma)$ reaction in novae} 

\author{E.\,C.~Vyfers}
\affiliation{Department of Physics and Astronomy, University of the Western Cape, P/B X17, Bellville 7535, South Africa}%

\author{V.~Pesudo}
\affiliation{Department of Physics and Astronomy, University of the Western Cape, P/B X17, Bellville 7535, South Africa}%
\affiliation{iThemba LABS, P.O. Box 722, Somerset West 7129, South Africa}

\author{S.~Triambak}
\email{striambak@uwc.ac.za}
\affiliation{Department of Physics and Astronomy, University of the Western Cape, P/B X17, Bellville 7535, South Africa}%

\author{M.~Kamil}
\affiliation{Department of Physics and Astronomy, University of the Western Cape, P/B X17, Bellville 7535, South Africa}%

\author{P.~Adsley}
\affiliation{iThemba LABS, P.O. Box 722, Somerset West 7129, South Africa}%
\affiliation{Cyclotron Institute and Department of Physics \& Astronomy, Texas A\&M University, College Station, Texas 77843, USA}%

\author{B.\, A.~Brown}
\affiliation{Department of Physics and Astronomy and the Facility for Rare Isotope Beams, Michigan State University, East Lansing, Michigan 48824-1321, USA}
\author{H.~Jivan}
\affiliation{School of Physics, University of the Witwatersrand, Johannesburg 2050, South Africa}%

\author{D.\,J.\,Marin-Lambarri}
\affiliation{Department of Physics and Astronomy, University of the Western Cape, P/B X17, Bellville 7535, South Africa}%
\affiliation{iThemba LABS, P.O. Box 722, Somerset West 7129, South Africa}
\affiliation{Instituto de F\'isica, Universidad Nacional Aut\'onoma de M\'exico, Av. Universidad 3000, Mexico City 04510, Mexico}


\author{R.~Neveling}
\affiliation{iThemba LABS, P.O. Box 722, Somerset West 7129, South Africa}%

\author{J.\,C.\,Nzobadila~Ondze}
\affiliation{Department of Physics and Astronomy, University of the Western Cape, P/B X17, Bellville 7535, South Africa}%

\author{P.~Papka}
\affiliation{Department of Physics, University of Stellenbosch, Private Bag X1, 7602 Matieland, Stellenbosch, South Africa}%

\author{L.~Pellegri}
\affiliation{iThemba LABS, P.O. Box 722, Somerset West 7129, South Africa}%
\affiliation{School of Physics, University of the Witwatersrand, Johannesburg 2050, South Africa}%

\author{B.\,M.~Rebeiro}
\affiliation{Department of Physics and Astronomy, University of the Western Cape, P/B X17, Bellville 7535, South Africa}%
\author{B.~Singh}
\affiliation{Department of Physics and Astronomy, University of the Western Cape, P/B X17, Bellville 7535, South Africa}%

\author{F.\,D.~Smit}
\affiliation{iThemba LABS, P.O. Box 722, Somerset West 7129, South Africa}%

\author{G.\,F.~Steyn}
\affiliation{iThemba LABS, P.O. Box 722, Somerset West 7129, South Africa}%

\date{\today}
%
%
%

\begin{abstract}
\noindent \textbf{Background:} The nucleosynthesis of several proton-rich nuclei is determined by radiative proton-capture reactions on unstable nuclei in nova explosions. One such reaction is \isotope[23]{Mg}$(p,\gamma)$\isotope[24]{Al}, which links the NeNa and MgAl cycles in oxygen-neon (ONe) novae.\\

\noindent \textbf{Purpose:} To extract \isotope[23]{Mg}$(p,\gamma)$ resonance strengths from a study of proton-unbound states in \isotope[24]{Al}, produced via the $^{24}$Mg($^{3}$He,$t$) reaction.\\

\noindent \textbf{Methods:} A beam of \isotope[3]{He}$^{2+}$ ions at 50.7~MeV was used to produce the states of interest in \isotope[24]{Al}. Proton-triton angular correlations were measured with a $K=600$ QDD magnetic spectrometer and a silicon detector array, located at iThemba LABS, South Africa.\\

\noindent \textbf{Results:} 
We measured the excitation energies of the four lowest proton-unbound states in $^{24}$Al  
and place lower-limits on $\Gamma_p/\Gamma$ values for these four states. Together with shell-model calculations of partial gamma widths, the experimental data are also used to determine resonance strengths for the four lowest \isotope[23]{Mg}$(p,\gamma)$\isotope[24]{Al} resonances. \\

\noindent \textbf{Conclusions:} 
The energy of the dominant first \isotope[23]{Mg}$(p,\gamma)$ resonance is determined to be $E_{r} = 478 \pm 4$~keV, with a resonance strength $\omega \gamma = 19 \pm 9$~meV. 
\\
\end{abstract}

\maketitle

\section{Introduction}
\begin{figure}
\centering
\includegraphics[scale=0.7]{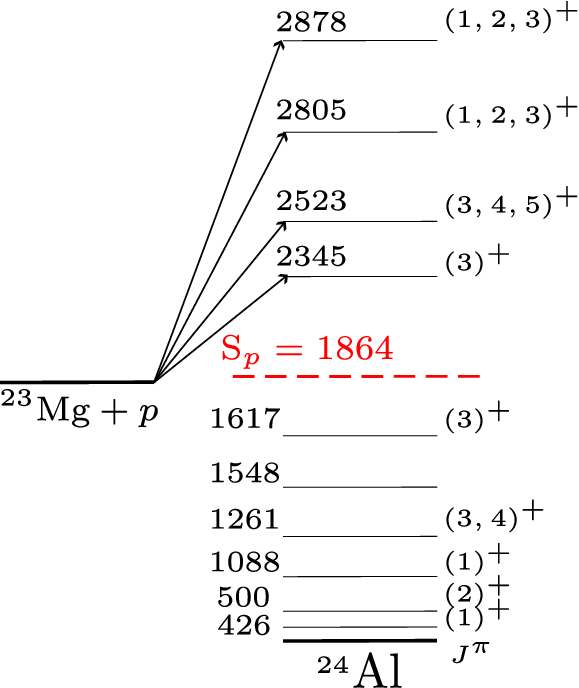} 
\caption{\isotope[23]{Mg}$(p,\gamma)$\isotope[24]Al resonances discussed in this work. Nominal energies (in keV) and spin-parity assignments are from Ref.~\cite{NDS}.}
\label{fig:Level_scheme}
\end{figure}

The \isotope[23]Mg$(p,\gamma)$\isotope[24]Al reaction is a crucial link between the NeNa and MgAl cycles in oxygen-neon (ONe) novae. Therefore, an accurate determination of this stellar reaction rate is critical for an improved understanding of elemental abundances up to Ca~\cite{Kelley:13,Jose:04}. At peak nova temperatures $(T_9 = 0.2-0.4)$, the dominant contribution to this process is through the first resonance above the proton threshold in $^{24}$Al  (c.f. Fig.~\ref{fig:Level_scheme}). 


Direct measurements of the $^{23}$Mg($p,\gamma)$ reaction rate are challenging, primarily because of small cross sections at astrophysically relevant energies and the difficulty in producing intense $^{23}$Mg radioactive ion beams with sufficient purity. As a result, there have been several attempts to determine the $^{23}$Mg($p,\gamma)$ reaction rate indirectly~\cite{Wallace,H_Herndl,M_Wiescher,S_Kubono,D_Visser}, using theoretical estimates of the partial proton and gamma widths of relevant states in \isotope[24]{Al}.
For isolated narrow resonances, the resonant contribution for each level can be obtained from its resonance strength
\begin{equation}
    \omega \gamma = \frac{(2J_r+1)}{8}\frac{\Gamma_p \Gamma_\gamma}{\Gamma},
\end{equation}
where $J_r$ is the total angular momentum of the resonant state, $\Gamma_p$ and $\Gamma_\gamma$ are its partial proton and gamma widths, and $\Gamma = \Gamma_p + \Gamma_\gamma$ is the total width.  

To date, the only direct \isotope[23]{Mg}$(p,\gamma)$ measurement was performed by Erikson \textit{et al.}~\cite{Erikson} at the TRIUMF-ISAC facility, where a radioactive $^{23}$Mg ion beam was incident on a hydrogen gas target at the DRAGON recoil spectrometer~\cite{dragon}. Prompt $\gamma$ rays were detected with an array of bismuth germanate (BGO) scintillators surrounding the target, while the recoils were identified using a combination of ionization chamber and microchannel plate (MCP) detectors located at the focal plane of the spectrometer. The experiment was severely affected by a dominant time-varying \isotope[23]Na contamination in the beam~\cite{Erikson}. 
Nevertheless,  the energy of the lowest resonance and its corresponding resonance strength were determined to be
$E_r = 485.7_{-1.8}^{+1.3}$~keV and $\omega \gamma = 37.8_{-15.4}^{+20.5}$~meV, respectively~\cite{Erikson}. The former translates to an excitation energy of $E_{\rm x} = 2349.8^{+1.3}_{-1.8}$~keV in $^{24}{\rm Al}$, a value that is in tension with an independent $\gamma$ ray measurement, $E_{\rm x} = 2345.1 \pm 1.4$~keV~\cite{G_Lotay}.


Taking the above into consideration, in this work we report complementary studies of proton-unbound states in \isotope[24]Al, produced via the \isotope[24]Mg$(^{3}{\rm He},t)$\isotope[24]Al charge-exchange reaction.  

\section{APPARATUS}
\begin{figure}[t]
\includegraphics[scale=0.3]{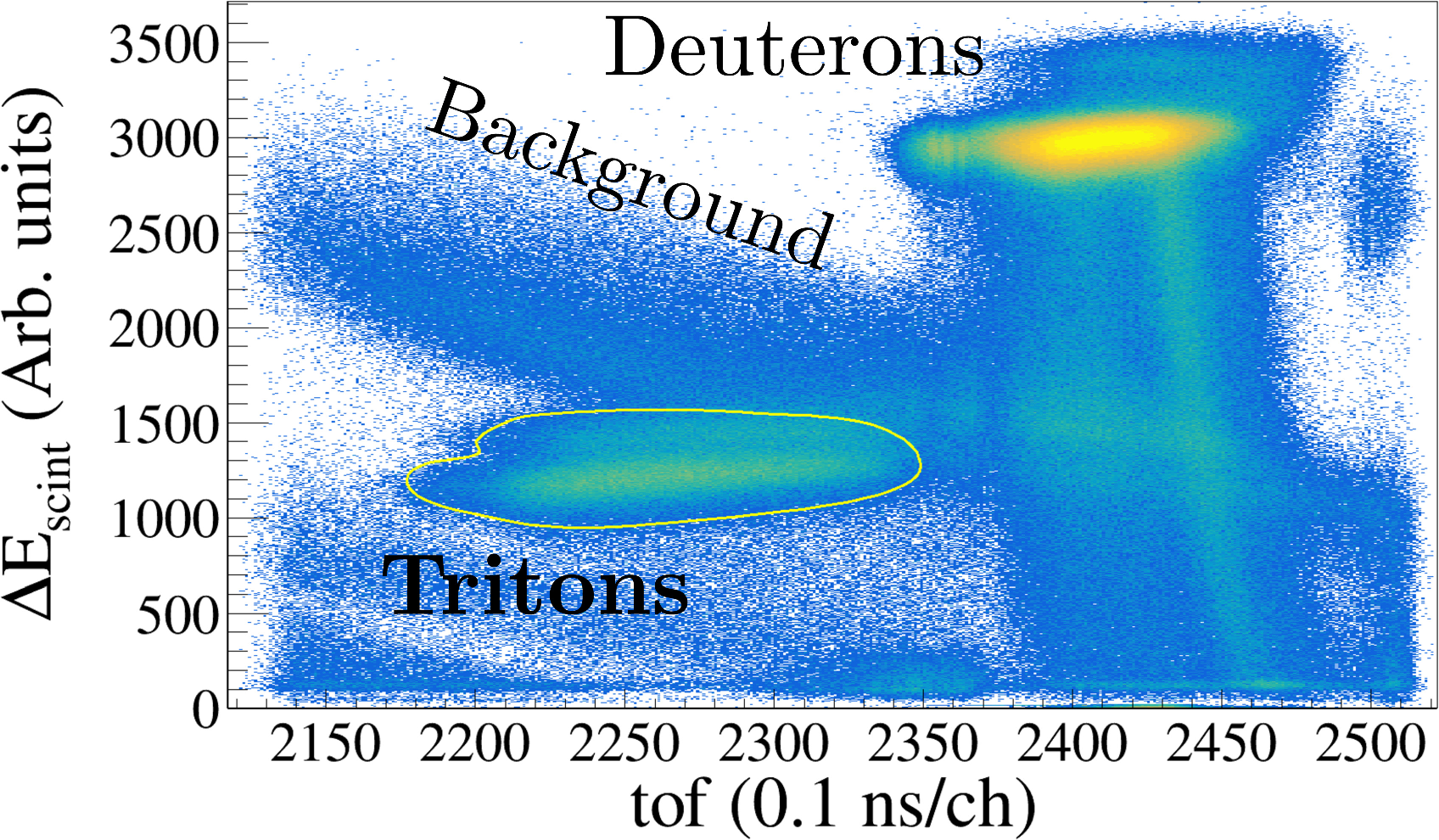} 
\caption{PID spectrum from energy deposited in the scintillator vs relative time of flight (tof), as discussed in the text. }
\label{fig:PID}
\end{figure}

The experiment was performed at the iThemba LABS cyclotron facility, where a 10~pnA, 50.7-MeV dispersion-matched $^{3}$He$^{2+}$ beam was bombarded on an $\approx300~\mu\text{g}/\text{cm}^{2}$-thick MgF$_{2}$ target on a carbon backing. The target was located in the scattering chamber of the $K = 600$ QDD magnetic spectrometer, which was configured in $0^\circ$ mode, with the beam stop located inside the first dipole magnet~\cite{k600}. Momentum-analyzed reaction ejectiles were detected in a $1/4^{\prime\prime}$-thick plastic scintillator at the spectrometer focal plane, after passing through two vertical drift chambers (VDCs). The VDCs determined the horizontal and vertical positions of the ejectiles crossing the focal plane, while the plastic scintillator was used for particle identification (PID) purposes, and to generate the data acquisition (DAQ) trigger. The PID plots were generated using the relative time difference between the cyclotron RF and DAQ trigger signals, together with the energy deposited in the scintillator. Fig.~\ref{fig:PID} shows the PID plot obtained for this experiment, with the triton group explicitly highlighted. 

An array of five 400-$\mu$m-thick MMM-type Double-sided Silicon Strip Detectors (DSSDs) called CAKE~\cite{Adsley_2017} was used to detect protons from unbound states in \isotope[24]Al, in coincidence with triton focal-plane events. The detectors were placed upstream of the target location in a backward-facing lampshade configuration and covered $\approx 25\%$ of the total solid-angle. Each DSSD comprised 16 ring channels and 8 sector channels. The rings had an angular range of $\theta_{lab} = 115 \degree - 165 \degree$ for the whole array.



\section{DATA ANALYSIS}
\subsection{Triton singles}
\begin{figure}
    \centering
    \includegraphics[scale=0.34]{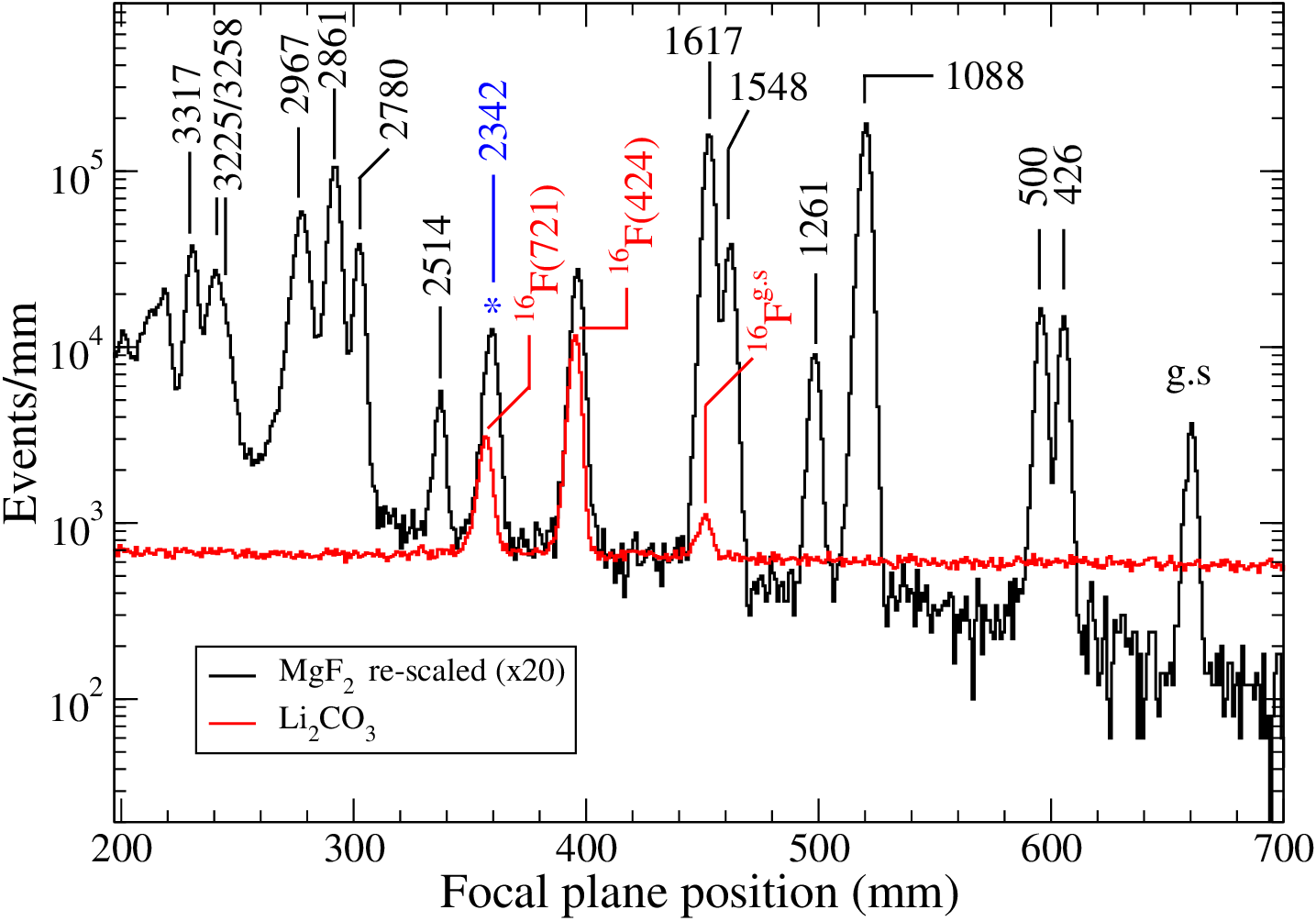}
    \caption{Triton focal-plane spectra obtained for the MgF$_{2}$ and Li$_{2}$CO$_{3}$ targets. Nominal $^{24}$Al energies are labeled (in keV), with the state corresponding to the first \isotope[23]{Mg}$(p,\gamma)$ resonance marked in blue. }
    \label{fig:triton_overlaid}
\end{figure}

The \isotope[24]{Mg}$(^3{{\rm He}},t)$ singles spectrum, obtained with the appropriate triton PID gate is shown in Fig.~\ref{fig:triton_overlaid}. The full widths at half-maximum (FWHM) for the triton peaks were $\approx 40$~keV.
Because the \isotope[24]{Mg}$(^3{{\rm He}},t)$ reaction $Q$-value is $\sim 10$~MeV lower than \isotope[25,26]{Mg}$(^3{{\rm He}},t)$ reactions, one can safely assume negligible contributions from other Mg isotopes in the spectrum. However other contaminant peaks from carbon and oxygen in the target foil cannot be ruled out. To determine such contributions, we took additional $(\isotope[3]{He},t)$ data with a Li$_{2}$CO$_{3}$ target, whose spectrum is also shown in Fig.~\ref{fig:triton_overlaid}. Three contaminant peaks are evident in the region of interest (ROI). We associate these peaks to states in \isotope[16]{F}, produced via the \isotope[16]{O}$(^3{{\rm He}},t)$ reaction.

All triton peaks were fit using a lineshape function defined by a Gaussian distribution convolved with a low-energy exponential tail~\cite{Triambak, Kamil}. Peak centroids $(\mu_t)$ and areas $(A_t)$ were recorded for later analysis. Known excitation energies in $^{24}$Al were next used to perform an \textit{in-situ} energy calibration of the focal-plane spectrum. This was done using a quadratic fit of the form
\begin{equation}
 E_{\rm fit}(i) = a_0 + a_1 \mu_t(i) + a_2 \mu_t(i)^2.  
\label{eq:calib}
\end{equation} 
For this procedure, we first referred to the Nuclear Data Sheets~(NDS)~\cite{NDS}, in order to identify calibration peaks within the range $425 \lesssim E_\mathrm{x} \lesssim 3300$~keV. As a precautionary measure, we excluded the levels located at 1.5 and 1.6~MeV from our analysis, mainly because of possible unresolved doublets in the region~\cite{NDS,D_Visser}, and the presence of the contaminant \isotope[16]{F} ground state peak. 
The other states in \isotope[24]Al that provide calibration peaks in the ROI are shown in Table~\ref{table:calib}.
\begin{table}[t]
\begin{flushleft}
\caption{States in \isotope[24]Al used for energy calibration}
\begin{ruledtabular}
\begin{tabular}{c c c}
\multicolumn{3}{c}{Excitation energy~(keV)}\\
\cline{1-3}
\multicolumn{1}{c}{Most precise value}& \multicolumn{1}{c}{From the NDS~\cite{NDS}}&\multicolumn{1}{c}{Adjusted value}\\
\colrule
\label{table:calib}
$425.8 \pm 0.1^{\text{a}}$ & $425.81 \pm 0.10$ & ...\\
$500.1 \pm 0.1^{\text{b}}$ & $500.12 \pm 0.13$ & ...\\
$1088.2 \pm 0.2^{\text{b}}$ & $1088.35 \pm 0.22$ & ...\\
$1261.2 \pm 0.3^{\text{b}}$ & $1261.09 \pm 0.22$ &...\\
$2978 \pm 6^{\text{c}}$ &$2978 \pm 6$ & $2967.0 \pm 4.4$ \\
$3236 \pm 6^{\text{c}}$ &$3236 \pm 6$ & $3225.0 \pm 4.4$ \\
$3269 \pm 6^{\text{c}}$ &$3269 \pm 6$ & $3258.0 \pm 4.4$ \\
\end{tabular}
\end{ruledtabular}
$^{\text{a}}$ From J.~Honkanen~{\it et al.}~\cite{Honkanen}.\\
$^{\text{b}}$ From G.~Lotay~{\it et al.}~\cite{G_Lotay}.\\
$^{\text{c}}$ From D.\,W.~Visser~{\it et al.}~\cite{D_Visser}. The $6$~keV uncertainty was obtained from combining a $\pm 3$~keV statistical uncertainty with uncertainty contributions from relative target thicknesses and relative $Q$ values, at $\pm 3$~keV and $\pm 4.1$~keV respectively. \\
\end{flushleft}
\end{table}
It is evident from the table that excitation energies up to $E_\mathrm{x} = 1261$~keV are known with reasonably high precision. This is because they were primarily determined via $\gamma$ ray measurements. In comparison, the energies of the higher-lying states have much larger uncertainties of $\pm 6$~keV. These values are from a previous \isotope[24]Mg$(^3{{\rm He}},t)$\isotope[24]{Al} measurement by Visser~{\it et~al.}~\cite{D_Visser}, who used the \isotope[28]Si$(^3{{\rm He}},t)$\isotope[28]P reaction for energy calibration. However, there has been a revised determination of the $^{28}$P mass-excess since then. The present atomic mass data compilation~\cite{ame2021} quotes it at $-7147.9 \pm 1.1$~keV, which is approximately 11~keV higher than the previous accepted value~\cite{masses:03}. Consequently, we adjusted the excitation energies of the higher-lying states based on this difference. The reduced uncertainties on the adjusted values are because of a smaller uncertainty contribution ($\pm 1.2$~keV) for the relative $Q$ values of the two reactions, as determined from the updated masses in Ref.~\cite{ame2021}.

The results of our regression analysis were used to determine the excitation energies of observed states in \isotope[24]{Al} that are relevant for the \isotope[23]{Mg}$(p,\gamma)$ reaction rate. These states are listed in Table~\ref{table:Ex_values}.\footnote{We do not see any explicit signature of the 3019~keV state reported by Ref.~\cite{D_Visser}. Our results and conclusions remain unaffected on excluding the 2.9~MeV state from the calibration procedure. Our data also do not show evidence of the recently reported state at 2605~keV~\cite{Zegers}.} As an example, for the lowest \isotope[23]{Mg}$(p,\gamma)$ resonance, our calibration yields $E_\mathrm{x} = 2342 \pm 4$~keV, which is in reasonable agreement with a previous high-precision $\gamma$ ray measurement~\cite{G_Lotay,NDS}. In order to correct for the contamination from $^{16}$O in the target, for this particular case the ROI was fitted with a doublet of peaks, with one peak centroid kept fixed at the value for the 721~keV state, which was obtained from the Li$_{2}$CO$_{3}$ $(^{3}{\rm He},t)$ spectrum.

It was also important to correct for the \isotope[16]{O}$(^3{{\rm He}},t)$\isotope[16]F contamination, when the area of the 2342~keV peak was used to determine its corresponding proton branching-ratio. However such a correction is trivial for this case, because of the additional high statistics triton peak associated with the 424~keV state in \isotope[16]{F}. Since this peak appears in both spectra (around channel number 390 in~Fig.~\ref{fig:triton_overlaid}), the correction factor was determined from the relative areas of both the 424 and 721~keV contaminant peaks.

%
\begin{table}[htpb]
\begin{flushleft}
\caption{Extracted energies for the four lowest \isotope[23]{Mg}$(p,\gamma)$ resonances.}
\begin{ruledtabular}
\begin{tabular}{c c c}
    \multicolumn{1}{c}{Previous work}&\multicolumn{2}{c}{This work}\\
\cline{2-3}
\multicolumn{1}{c}{$E_\mathrm{x}$ (keV)}&\multicolumn{1}{c}{$E_\mathrm{x}$ (keV)} &\multicolumn{1}{c}{$E_r$ (keV)}\\
\colrule
\label{table:Ex_values}
$2345.1 \pm 1.4^{\text{a}}$ & $2342 \pm 4^{\text{e}}$ & $478 \pm 4$\\
$2349.8^{+1.3}_{-1.8}$$^{\text{b}}$ & & \\
$2523 \pm 3^{\text{c}}$ & $2513 \pm 4$ & $649 \pm 4$\\
$2805 \pm 10^{\text{d}}$ & $2779 \pm 4$ & $915 \pm 4$\\
$2878 \pm 6^{\text{d}}$ & $2860 \pm 5$ & $996 \pm 5$
\end{tabular}
\end{ruledtabular}
$^{\text{a}}$ From G.~Lotay~{\it et al.}~\cite{G_Lotay}.\\
$^{\text{b}}$ From the resonance energy reported by Erikson \textit{et al.}~\cite{Erikson}.\\
$^{\text{c}}$ From D.\,W.~Visser~{\it et al.}~\cite{Visser2}.\\
$^{\text{d}}$ From Ref.~\cite{NDS}.\\
$^{\text{e}}$ The centroid for this peak in the singles spectrum is consistent with the one obtained with the proton-gated triton coincidences, described in Section~\ref{sect:3.2}. The latter yield an excitation energy of $2340 \pm 5$~keV.
\end{flushleft}
\end{table}
\subsection{Triton-proton coincidences}
\label{sect:3.2}
As mentioned previously, charged-particle-like events from \isotope[24]{Mg}$(\isotope[3]{He},t)$\isotope[24]{Al}$^*(p)$\isotope[23]{Mg} were detected using the DSSD array described in Ref.~\cite{Adsley_2017}. 
The energy and timing signals for the array were digitized using CAEN V785 ADCs and V1190A TDCs, respectively.   Triton-proton ($t$-$p$) coincidences were selected by gating on the prompt TDC timing peak, which was around 40~ns wide and further imposing an energy condition on the sector and ring channels (so that they are within 300~keV). The DSSDs were energy calibrated using a \isotope[226]{Ra} $\alpha$-source at a later time, when the beam was off.
\begin{figure}[t]
\includegraphics[scale=0.32]{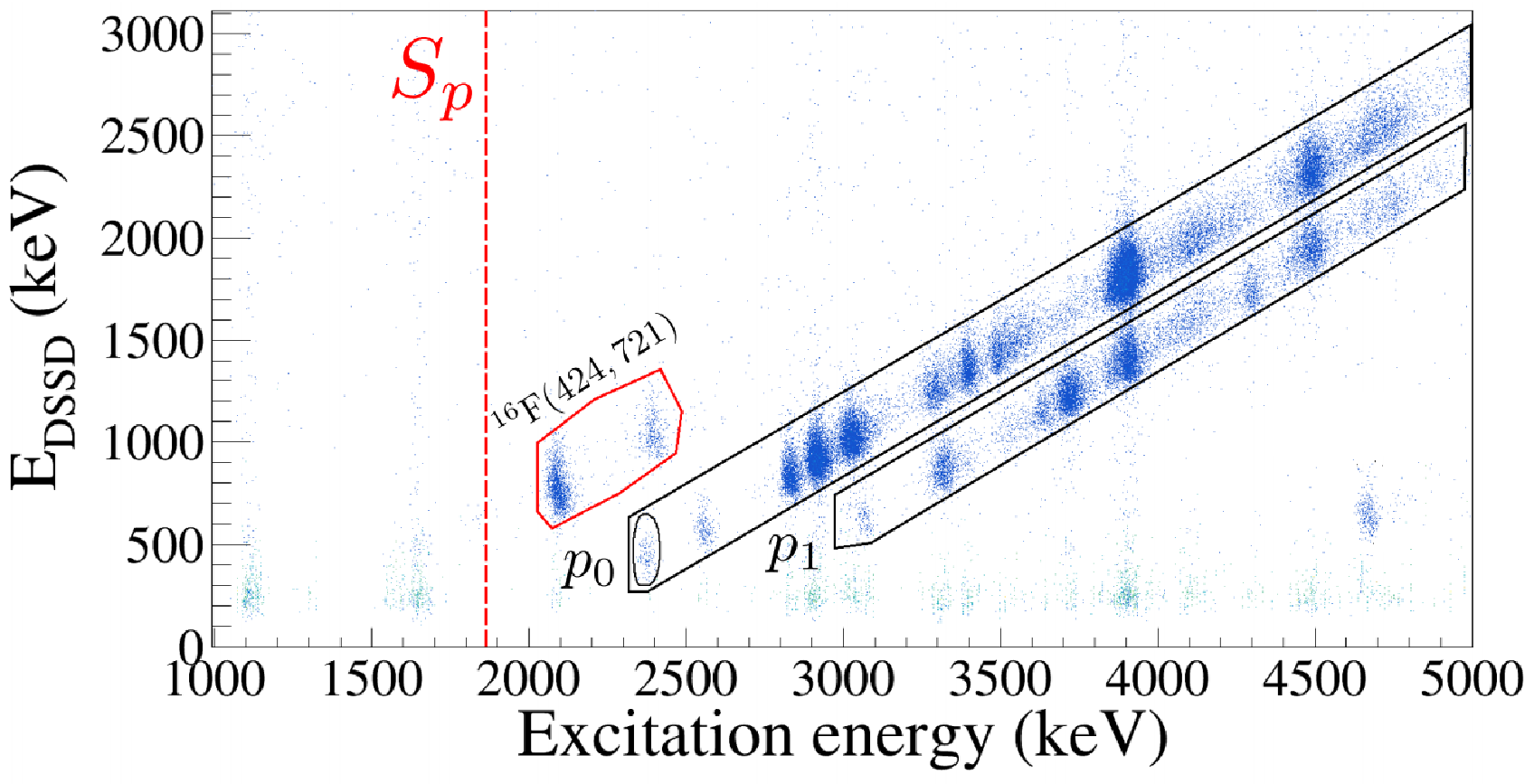} 
\caption{$t$-$p$ coincidence data, with proton groups from states in $^{24}$Al highlighted by diagonal bands. The upper (\textit{$p_{0}$}) and lower (\textit{$p_{1}$}) bands are from transitions to the ground and first excited states in $^{23}$Mg, respectively. The ellipse around the $p_{0}$ proton group is for the lowest $^{23}$Mg($p$,$\gamma$)$^{24}$Al resonance.}
\label{fig:Coincidence_data}
\end{figure}

Fig.~\ref{fig:Coincidence_data} shows the coincident $t$-$p$ energy loci obtained from this experiment. Since each ring of the DSSD array corresponds to a polar angle $(\theta)$ with respect to the beam-axis, the proton branching ratios $B_p(\theta)$ can be determined by gating on the relevant proton groups in Fig.~\ref{fig:Coincidence_data}, and making use of the formula 
\begin{equation}
\label{eq:branch}
B_p(\theta) = \frac{N_{tp}(\theta)}{N_t} \frac{1}{\epsilon(\theta)}.
\end{equation}
Here $N_{tp}(\theta)$ are the registered $t$-$p$ coincidences for a particular ring, $N_t$ are the corresponding triton singles and $\epsilon(\theta)$ is the proton detection efficiency of the ring. For states with definite spin and parity, these yields are expected to follow the simple angular distribution~\cite{Noe:1974} 
\begin{equation}
\label{eq:Leg_Pol}
B_p(\theta_\mathrm{c.m.}) = \sum_{k=\text{even}} A_{k} P_{k}(\cos\theta_\mathrm{c.m.}),
\end{equation}
whose integrated yield gives the total proton branching ratio for each state  
\begin{equation}
 \frac{\Gamma_p} {\Gamma} = \int_{-1}^{1}B_p(\theta_\mathrm{c.m.}) d(\cos \theta_\mathrm{c.m.}) = A_0.
\end{equation}
 
\begin{figure}[t]
\centering{\includegraphics[scale=0.32]{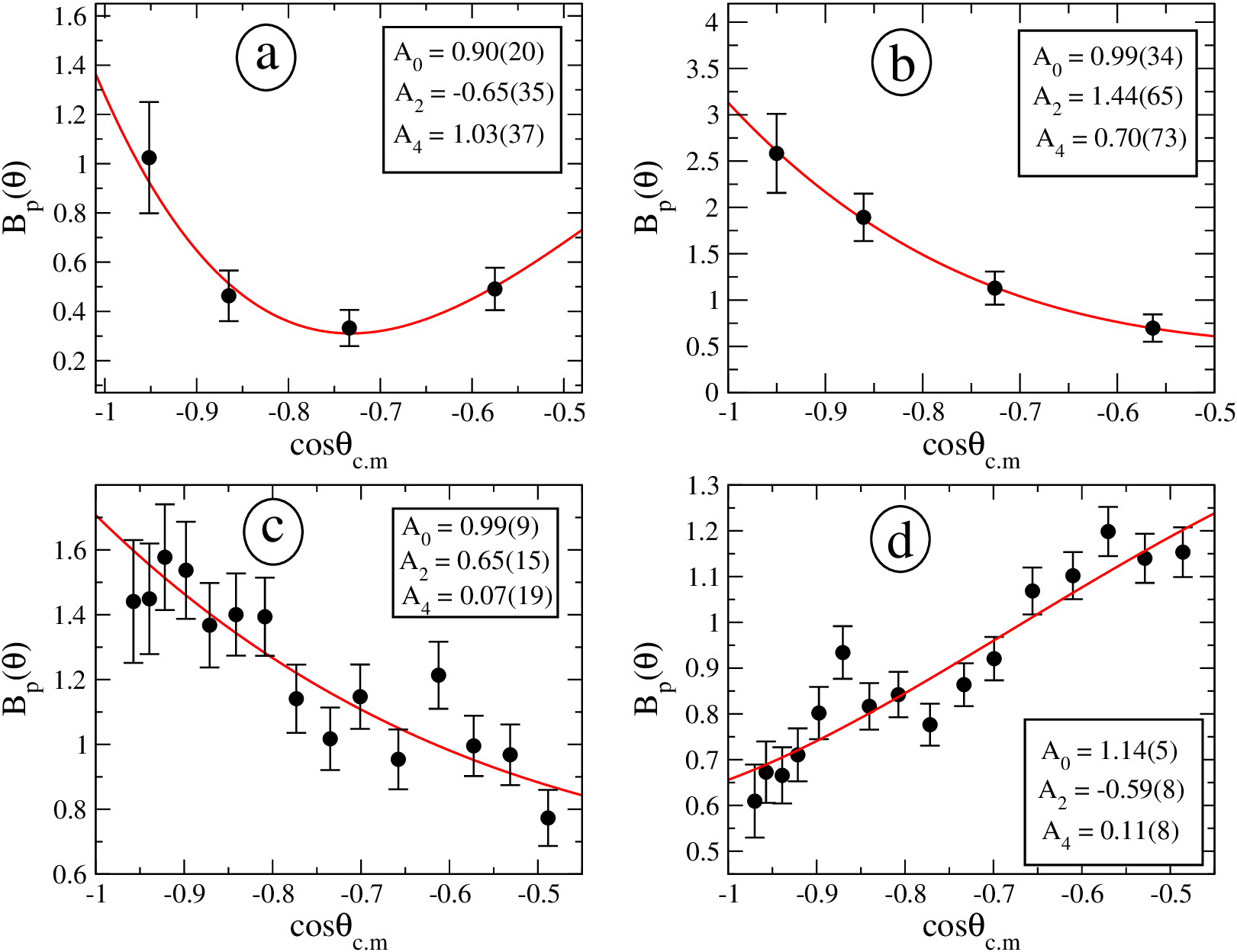}}
\caption{Legendre polynomial fits to the measured proton-triton angular distributions. The states in $^{24}$Al are labeled by their nominal energies \textbf{a}: $E_{\rm x} = 2342$~keV, \textbf{b}: $E_{\rm x} = 2513$~keV, \mbox{\textbf{c}: $E_{\rm x} = 2779$~keV and \textbf{d}: $E_{\rm x} = 2860$~keV.}}
\label{fig:AngDistRES_multi} 
\end{figure}

To perform the above analysis, we first determined $\epsilon(\theta)$ for each active ring of the DSSD array using GEANT4 Monte Carlo simulations. These values were used to obtain an initial set of $B_p(\theta)$ for the first four \isotope[23]{Mg}$(p,\gamma)$\isotope[24]{Al} resonances. Next, a similar analysis was also performed on the \isotope[16]{O}$(^3{{\rm He}},t)\isotope[16]{F}^*(p)$ data, obtained with the Li$_{2}$CO$_{3}$ target. As \isotope[16]{F} is unbound and $\Gamma_p/\Gamma = 1$ for its observed states~\cite{NNDC,Lee:2007}, this important procedure determined an effective normalization\footnote{A weighted mean of the results from the 424 and 721~keV-state data yielded a normalization factor $\kappa = 1.51 \pm 0.04$.} to correct the \isotope[24]{Al} results, which had previously only relied on simulated efficiencies. 
The renormalized $t$-$p$ distributions for the first four proton-unbound states in \isotope[24]{Al} are shown in Fig.~\ref{fig:AngDistRES_multi}.\footnote{As the coincidence data for the 2342 and 2513~keV states were statistics limited, for these cases the angular yield at each bin was obtained by combining data from 4 adjacent strips in the DSSDs.}, with the converged fit results for $A_0$ and their associated $\pm 1\sigma$ uncertainties shown in the figure insets. If one assumes Gaussian distributions for these values, it is apparent that their coverage probabilities also include disallowed regions of parameter space, with $\Gamma_p/\Gamma > 1$. Consequently, we use the fit-results in Fig.~\ref{fig:AngDistRES_multi} to present lower-limits on $\Gamma_p/\Gamma$ values for all four states. These are listed in Table~\ref{table:Res_pars}.

\section{RESULTS}
Although we are unable to quote $\Gamma_p/\Gamma$ values with sufficient precision, it is still possible to extract meaningful resonance strengths from our data. 
For this part of the analysis, we performed shell model calculations of partial gamma widths ($\Gamma_\gamma$), using the isospin non-conserving (INC) Hamiltonian of Ormand and Brown (OB)~\cite{Ormand_Brown} and the more recently developed USD-C Hamiltonian~\cite{usdc}. These values are listed in Table~\ref{table:Res_pars}. For comparsion we also show the results of a previous calculation by Herndl~{\it et al.}~\cite{H_Herndl}. Our results indicate that the theoretical uncertainty in the calculated $\Gamma_\gamma$ values are about 20-50\%. We also performed calculations of level lifetimes in the mirror \isotope[24]{Na} nucleus. Table~\ref{table:theory} shows a comparison of these results, with experimentally measured values~\cite{NDS} for states above 2~MeV. For these theory calculations, effective charges and $g$-factors for the $E2$ and $M1$ operators were taken from the global sd-shell fit results in Ref.~\cite{Richter:08}.

\begin{table}[t]
\begin{flushleft}
\caption{Parameters for key $^{23}$Mg($p$,$\gamma$)$^{24}$Al resonance reactions.}
\begin{ruledtabular}
\begin{tabular}{c c c c c c c}
$J_n^\pi$ & $E_{r}$ & $\Gamma_{p}/\Gamma~^\text{a}$  & \multicolumn{3}{c}{$\Gamma_\gamma$~(meV)}& $\omega \gamma$\\
\cline{4-6}
Ref.~\cite{H_Herndl}& (keV) & & Ref.~\cite{H_Herndl} & OB$^{\text{b}}$& USD-C$^{\text{c}}$ & (meV)\\
\colrule
$3_3^+$ & $478 \pm 4$	&$\ge 0.64$ & 33 & 24 & 23 & $19 \pm 9$ \\
$4_2^+$ & $649 \pm 4$	&$\ge 0.55$     & 53 & 45 & 39 & $ 50 \pm 25$ \\
$2_4^+$ & $915 \pm 4$	&$\ge 0.87$     & 83 & 108 & 71 &$ 67 \pm 34$\\
$3_4^+$ & $996 \pm 5$ &$\ge 0.99^{\text{d}}$& 14 & 12 & 8 & $11 \pm 5$
\label{table:Res_pars}
\end{tabular}
\end{ruledtabular}
$^{\text{a}}$Unless specified otherwise, lower limits are quoted at the 90\% confidence level (CL).\\
$^{\text{b}}$Evaluated using the INC Hamiltonian from Ormand and Brown~\cite{Ormand_Brown}.\\
$^{\text{c}}$Evaluated using the USD-C INC Hamiltonian from Ref.~\cite{usdc}.\\
$^{\text{d}}$ Quoted at the 99.9\% CL. \\
\end{flushleft}
\end{table}

As the $^{23}{\rm Mg}(p,\gamma)$ cross section is dominated by the $3_3^+$ level in \isotope[24]Al, we note that the Ormand-Brown result for the analog state in \isotope[24]Na is in better agreement with experiment. Consequently, we choose to determine $^{23}{\rm Mg}(p,\gamma)$ resonance strengths from the Ormand-Brown results for $\Gamma_\gamma$ in Table~\ref{table:Res_pars}, each with a conservatively-assigned 50\% relative uncertainty.

\begin{table}[htpb]
\begin{flushleft}
\caption{Half-lives of states in the mirror \isotope[24]{Na}. A comparison between theory and experiment.}
\begin{ruledtabular}
\begin{tabular}{c c c c c}
Energy$^\text{a}$  & $J_n^\pi$ & \multicolumn{3}{c}{$T_{1/2}$~(fs)}\\
\cline{3-5}
(keV) &  & OB& USD-C & Expt$^\text{a}$\\
\colrule
$2513$ & $3_3^+$	          &$17$  &23  &$10 \pm 5$\\
$2563$ & $4_2^+$	          &$10$  &10  &$< 17$\\
$2904$ & $3_4^+$	            &$19$  &38  &$35 \pm 8$\\
$2978$ & $2_4^+$	           &$4$   &6.5  &$< 17$\\
$3216$ & $4_3^+$	            &$15$  &12  &$16 \pm 6$
\label{table:theory}
\end{tabular}
\end{ruledtabular}
$^{\text{a}}$ From Ref.~\cite{NDS}. \\
\end{flushleft}
\end{table}

Next, we used Monte Carlo simulations to generate Gaussianly distributed random deviates around the measured central $A_0$ values that correspond to individual resonances. Each simulated variable $A_0^\text{sim}$ was used to calculate a corresponding $\chi^2$ with respect to the measured data points in Fig.~\ref{fig:AngDistRES_multi}. Simultaneously, the simulations also determined associated $\omega \gamma$ values, making use of the relation
\begin{equation}
\left(\frac{\Gamma_p}{\Gamma_\gamma}\right)_\text{sim} = \frac{A_0^\text{sim}}{1-A_0^\text{sim}}. 
\end{equation}
For each state, the simulated $\omega \gamma$ that yielded the minimum $\chi^2$ with respect to our data\footnote{Events that yielded $\Gamma_p/\Gamma_\gamma < 0$ were rejected from the analysis.} was accepted as the true value, with its $\pm 68\%$ CL statistical uncertainties obtained from the values that yield $\chi^2 = \chi^2_\text{min}+1$. Since $\omega \gamma$ is proportional to $\Gamma_\gamma$, the 50\% (dominant) systematic uncertainty contributions from theory were added in quadrature to obtain the final uncertainties in the $\omega \gamma$ values, listed in Table~\ref{table:Res_pars}. 

Our extracted $\omega \gamma$ for the lowest dominant \isotope[23]Mg$(p,\gamma)$ resonance is $19 \pm 9$~meV. This may be compared with the direct-measurement result, $\omega \gamma = 37.8_{-15.4}^{+20.5}$~meV~\cite{Erikson}, which relied on a joint likelihood analysis to determine the median and $\pm 68\%$ CL values for the resonance parameters. In this previous study, the quoted uncertainties were the result of a highly asymmetric joint probability density function (PDF) in the $(E_r,\omega\gamma)$ parameter space, which showed significant tailing and a bimodal double-humped structure for its projected $\omega \gamma$ probability distribution. Although our extracted resonance strength does not significantly disagree with the result from Ref.~\cite{Erikson}, it appears to be more consistent with the left-peaked structure of their PDF for $\omega \gamma$ (c.f. Fig.~16 in Ref.~\cite{Erikson}), which is centered at approximately $23$~meV instead. This is consistent with independent observations by Wrede~{\it et al.}~\cite{Wrede}, who used revised data to reduce the bimodal joint PDF distribution from Ref.~\cite{Erikson} to a nearly unimodal one, with a new recommended median value $\omega \gamma = 26.6_{-7.0}^{+15.4}$~meV.
\begin{figure}[t]
\centering{\includegraphics[scale=0.35]{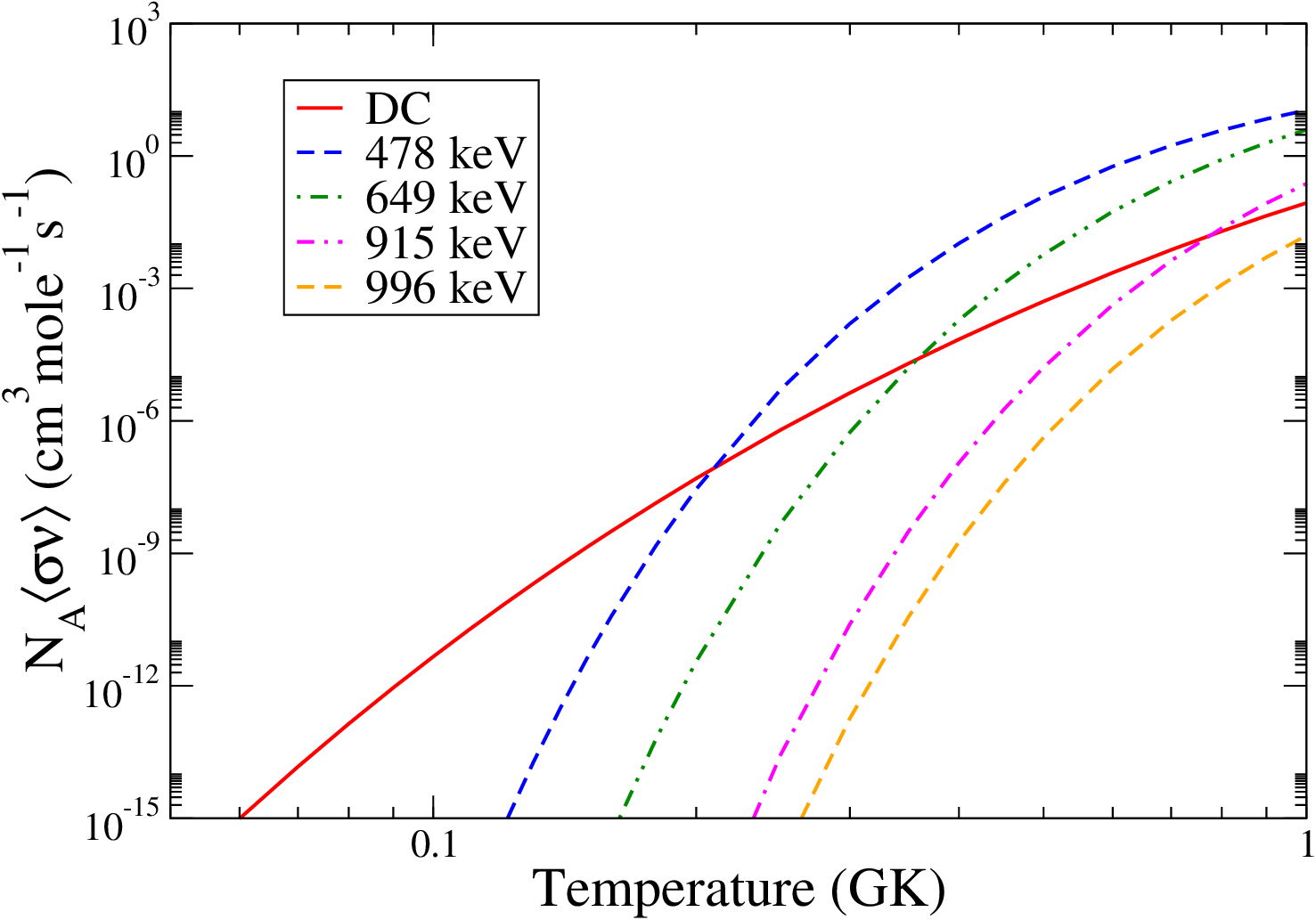}}
\caption{Median $^{23}$Mg($p$,$\gamma$)$^{24}$Al reaction rates obtained from this work. Direct capture (DC) and resonant contributions are shown separately, with the latter evaluated using the results listed in Table ~\ref{table:Res_pars}.}
\label{fig:stellar_rate_PAR}
\end{figure} 
\begin{figure}[t]
\centering{\includegraphics[scale=0.51]{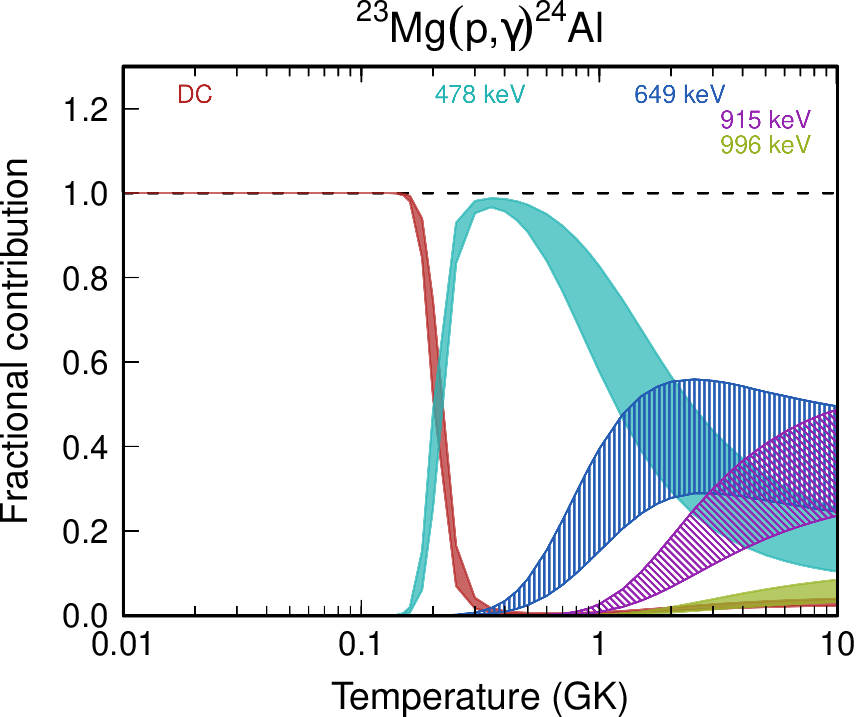
}}
\caption{Fractional contributions of direct capture (DC) and low-lying resonances to the total $^{23}$Mg($p$,$\gamma$)$^{24}$Al reaction rate. The uncertainty bands are evaluated using the Monte Carlo prescription described in Refs.~\cite{RatesMC1,RatesMC2}.}
\label{fig:Stellar_Contr} 
\end{figure} 

Finally, we used the values from Table~\ref{table:Res_pars} to evaluate \isotope[23]{Mg}$(p,\gamma)$ reaction rates with the RatesMC Monte Carlo code, described in Refs.~\cite{RatesMC1,RatesMC2}. The non-resonant (direct capture) contribution was determined from a polynomial expansion of the $S$-factor
\begin{equation}
 S(E) \approx S(0) + S'(0)E + \frac{1}{2}S''(0)E^2~\mathrm{keV~b},  
\end{equation}
whose coefficients were taken from Ref.~\cite{H_Herndl}. The results for individual resonances identified in this work are shown in Fig.~\ref{fig:stellar_rate_PAR}, with the fractional contribution of each component shown in Fig.~\ref{fig:Stellar_Contr}.

\section{Summary}
To summarize, this work reports a study of states in \isotope[24]{Al}, which are important to determine the \isotope[23]{Mg}$(p,\gamma)$\isotope[24]{Al} nuclear reaction rate in ONe novae. Our measured excitation energy for the dominant lowest resonance agrees with a previous higher-precision $\gamma$-ray measurement~\cite{G_Lotay}. Together with shell-model calculations of partial gamma widths, our experimental data are also used to extract key \isotope[23]{Mg}$(p,\gamma)$ resonance strengths. For the lowest resonance, we determine a resonance strength that is in reasonable with a previous direct measurement~\cite{Erikson}. It is noted however, that our extracted resonance strength is lower than the median value reported Ref.~\cite{Erikson}. This is consistent with results from other independent studies~\cite{Wrede,H_Herndl,S_Kubono}. Similarly, it can also be argued that at face-value, our results favor the energy reported by Lotay~{\it et al.}~\cite{G_Lotay}, for the first  \isotope[23]{Mg}$(p,\gamma)$\isotope[24]{Al} resonance.

\begin{acknowledgments}
We are thankful to Alejandro Garc\'ia for insightful discussions and Chris Wrede for directing us to Refs.~\cite{Visser2,Wrede}. This work was partially funded by the National Research Foundation (NRF), South Africa under Grant No. 85100. B.A.B acknowledges funding support from the National Science Foundation, under Grant. No.~PHY-2110365. E.C.V thanks the NRF-funded MaNuS/MatSci program at the University of the Western Cape for financial support during the course of his M.Sc.
\end{acknowledgments}
\bibliography{24al_ver3}

\begin{thebibliography}{29}%
\makeatletter
\providecommand \@ifxundefined [1]{%
 \@ifx{#1\undefined}
}%
\providecommand \@ifnum [1]{%
 \ifnum #1\expandafter \@firstoftwo
 \else \expandafter \@secondoftwo
 \fi
}%
\providecommand \@ifx [1]{%
 \ifx #1\expandafter \@firstoftwo
 \else \expandafter \@secondoftwo
 \fi
}%
\providecommand \natexlab [1]{#1}%
\providecommand \enquote  [1]{``#1''}%
\providecommand \bibnamefont  [1]{#1}%
\providecommand \bibfnamefont [1]{#1}%
\providecommand \citenamefont [1]{#1}%
\providecommand \href@noop [0]{\@secondoftwo}%
\providecommand \href [0]{\begingroup \@sanitize@url \@href}%
\providecommand \@href[1]{\@@startlink{#1}\@@href}%
\providecommand \@@href[1]{\endgroup#1\@@endlink}%
\providecommand \@sanitize@url [0]{\catcode `\\12\catcode `\$12\catcode
  `\&12\catcode `\#12\catcode `\^12\catcode `\_12\catcode `\%12\relax}%
\providecommand \@@startlink[1]{}%
\providecommand \@@endlink[0]{}%
\providecommand \url  [0]{\begingroup\@sanitize@url \@url }%
\providecommand \@url [1]{\endgroup\@href {#1}{\urlprefix }}%
\providecommand \urlprefix  [0]{URL }%
\providecommand \Eprint [0]{\href }%
\providecommand \doibase [0]{http://dx.doi.org/}%
\providecommand \selectlanguage [0]{\@gobble}%
\providecommand \bibinfo  [0]{\@secondoftwo}%
\providecommand \bibfield  [0]{\@secondoftwo}%
\providecommand \translation [1]{[#1]}%
\providecommand \BibitemOpen [0]{}%
\providecommand \bibitemStop [0]{}%
\providecommand \bibitemNoStop [0]{.\EOS\space}%
\providecommand \EOS [0]{\spacefactor3000\relax}%
\providecommand \BibitemShut  [1]{\csname bibitem#1\endcsname}%
\let\auto@bib@innerbib\@empty
\bibitem [{\citenamefont {Kelly}\ \emph {et~al.}(2013)\citenamefont {Kelly},
  \citenamefont {Iliadis}, \citenamefont {Downen}, \citenamefont {José},\ and\
  \citenamefont {Champagne}}]{Kelley:13}%
  \BibitemOpen
  \bibfield  {author} {\bibinfo {author} {\bibfnamefont {K.~J.}\ \bibnamefont
  {Kelly}}, \bibinfo {author} {\bibfnamefont {C.}~\bibnamefont {Iliadis}},
  \bibinfo {author} {\bibfnamefont {L.}~\bibnamefont {Downen}}, \bibinfo
  {author} {\bibfnamefont {J.}~\bibnamefont {José}}, \ and\ \bibinfo {author}
  {\bibfnamefont {A.}~\bibnamefont {Champagne}},\ }\href {\doibase
  10.1088/0004-637X/777/2/130} {\bibfield  {journal} {\bibinfo  {journal}
  {Astrophys. J.}\ }\textbf {\bibinfo {volume} {777}},\ \bibinfo {pages} {130}
  (\bibinfo {year} {2013})}\BibitemShut {NoStop}%
\bibitem [{\citenamefont {Jos\'e}\ \emph {et~al.}(2004)\citenamefont {Jos\'e},
  \citenamefont {Hernanz}, \citenamefont {Amari}, \citenamefont {Lodders},\
  and\ \citenamefont {Zinner}}]{Jose:04}%
  \BibitemOpen
  \bibfield  {author} {\bibinfo {author} {\bibfnamefont {J.}~\bibnamefont
  {Jos\'e}}, \bibinfo {author} {\bibfnamefont {M.}~\bibnamefont {Hernanz}},
  \bibinfo {author} {\bibfnamefont {S.}~\bibnamefont {Amari}}, \bibinfo
  {author} {\bibfnamefont {K.}~\bibnamefont {Lodders}}, \ and\ \bibinfo
  {author} {\bibfnamefont {E.}~\bibnamefont {Zinner}},\ }\href {\doibase
  10.1086/422569} {\bibfield  {journal} {\bibinfo  {journal} {Astrophys. J.}\
  }\textbf {\bibinfo {volume} {612}},\ \bibinfo {pages} {414} (\bibinfo {year}
  {2004})}\BibitemShut {NoStop}%
\bibitem [{\citenamefont {Wallace}\ and\ \citenamefont
  {Woosley}(1981)}]{Wallace}%
  \BibitemOpen
  \bibfield  {author} {\bibinfo {author} {\bibfnamefont {R.}~\bibnamefont
  {Wallace}}\ and\ \bibinfo {author} {\bibfnamefont {S.~E.}\ \bibnamefont
  {Woosley}},\ }\href@noop {} {\bibfield  {journal} {\bibinfo  {journal} {The
  Astrophysical Journal Supplement Series}\ }\textbf {\bibinfo {volume} {45}},\
  \bibinfo {pages} {389} (\bibinfo {year} {1981})}\BibitemShut {NoStop}%
\bibitem [{\citenamefont {Herndl}\ \emph {et~al.}(1998)\citenamefont {Herndl},
  \citenamefont {Fantini}, \citenamefont {Iliadis}, \citenamefont {Endt},\ and\
  \citenamefont {Oberhummer}}]{H_Herndl}%
  \BibitemOpen
  \bibfield  {author} {\bibinfo {author} {\bibfnamefont {H.}~\bibnamefont
  {Herndl}}, \bibinfo {author} {\bibfnamefont {M.}~\bibnamefont {Fantini}},
  \bibinfo {author} {\bibfnamefont {C.}~\bibnamefont {Iliadis}}, \bibinfo
  {author} {\bibfnamefont {P.~M.}\ \bibnamefont {Endt}}, \ and\ \bibinfo
  {author} {\bibfnamefont {H.}~\bibnamefont {Oberhummer}},\ }\href {\doibase
  10.1103/PhysRevC.58.1798} {\bibfield  {journal} {\bibinfo  {journal} {Phys.
  Rev. C}\ }\textbf {\bibinfo {volume} {58}},\ \bibinfo {pages} {1798}
  (\bibinfo {year} {1998})}\BibitemShut {NoStop}%
\bibitem [{\citenamefont {Wiescher}\ \emph {et~al.}(1986)\citenamefont
  {Wiescher}, \citenamefont {Gorres}, \citenamefont {Thielemann},\ and\
  \citenamefont {Ritter}}]{M_Wiescher}%
  \BibitemOpen
  \bibfield  {author} {\bibinfo {author} {\bibfnamefont {M.}~\bibnamefont
  {Wiescher}}, \bibinfo {author} {\bibfnamefont {J.}~\bibnamefont {Gorres}},
  \bibinfo {author} {\bibfnamefont {F.-K.}\ \bibnamefont {Thielemann}}, \ and\
  \bibinfo {author} {\bibfnamefont {H.}~\bibnamefont {Ritter}},\ }\href@noop {}
  {\bibfield  {journal} {\bibinfo  {journal} {Astronomy and Astrophysics}\
  }\textbf {\bibinfo {volume} {160}},\ \bibinfo {pages} {56} (\bibinfo {year}
  {1986})}\BibitemShut {NoStop}%
\bibitem [{\citenamefont {Kubono}\ \emph {et~al.}(1995)\citenamefont {Kubono},
  \citenamefont {Kajino},\ and\ \citenamefont {Kato}}]{S_Kubono}%
  \BibitemOpen
  \bibfield  {author} {\bibinfo {author} {\bibfnamefont {S.}~\bibnamefont
  {Kubono}}, \bibinfo {author} {\bibfnamefont {T.}~\bibnamefont {Kajino}}, \
  and\ \bibinfo {author} {\bibfnamefont {S.}~\bibnamefont {Kato}},\ }\href
  {\doibase https://doi.org/10.1016/0375-9474(95)00043-Z} {\bibfield  {journal}
  {\bibinfo  {journal} {Nuclear Physics A}\ }\textbf {\bibinfo {volume}
  {588}},\ \bibinfo {pages} {521} (\bibinfo {year} {1995})}\BibitemShut
  {NoStop}%
\bibitem [{\citenamefont {Visser}\ \emph {et~al.}(2007)\citenamefont {Visser},
  \citenamefont {Wrede}, \citenamefont {Caggiano}, \citenamefont {Clark},
  \citenamefont {Deibel}, \citenamefont {Lewis}, \citenamefont {Parikh},\ and\
  \citenamefont {Parker}}]{D_Visser}%
  \BibitemOpen
  \bibfield  {author} {\bibinfo {author} {\bibfnamefont {D.~W.}\ \bibnamefont
  {Visser}}, \bibinfo {author} {\bibfnamefont {C.}~\bibnamefont {Wrede}},
  \bibinfo {author} {\bibfnamefont {J.~A.}\ \bibnamefont {Caggiano}}, \bibinfo
  {author} {\bibfnamefont {J.~A.}\ \bibnamefont {Clark}}, \bibinfo {author}
  {\bibfnamefont {C.}~\bibnamefont {Deibel}}, \bibinfo {author} {\bibfnamefont
  {R.}~\bibnamefont {Lewis}}, \bibinfo {author} {\bibfnamefont
  {A.}~\bibnamefont {Parikh}}, \ and\ \bibinfo {author} {\bibfnamefont {P.~D.}\
  \bibnamefont {Parker}},\ }\href {\doibase 10.1103/PhysRevC.76.065803}
  {\bibfield  {journal} {\bibinfo  {journal} {Phys. Rev. C}\ }\textbf {\bibinfo
  {volume} {76}},\ \bibinfo {pages} {065803} (\bibinfo {year}
  {2007})}\BibitemShut {NoStop}%
\bibitem [{\citenamefont {Basunia}\ and\ \citenamefont
  {Chakraborty}(2022)}]{NDS}%
  \BibitemOpen
  \bibfield  {author} {\bibinfo {author} {\bibfnamefont {M.~S.}\ \bibnamefont
  {Basunia}}\ and\ \bibinfo {author} {\bibfnamefont {A.}~\bibnamefont
  {Chakraborty}},\ }\href {\doibase https://doi.org/10.1016/j.nds.2022.11.002}
  {\bibfield  {journal} {\bibinfo  {journal} {Nuclear Data Sheets}\ }\textbf
  {\bibinfo {volume} {186}},\ \bibinfo {pages} {3} (\bibinfo {year}
  {2022})}\BibitemShut {NoStop}%
\bibitem [{\citenamefont {Erikson}\ \emph {et~al.}(2010)\citenamefont
  {Erikson}, \citenamefont {Ruiz}, \citenamefont {Ames}, \citenamefont
  {Bricault}, \citenamefont {Buchmann}, \citenamefont {Chen}, \citenamefont
  {Chen}, \citenamefont {Dare}, \citenamefont {Davids}, \citenamefont {Davis},
  \citenamefont {Deibel}, \citenamefont {Dombsky}, \citenamefont {Foubister},
  \citenamefont {Galinski}, \citenamefont {Greife}, \citenamefont {Hager},
  \citenamefont {Hussein}, \citenamefont {Hutcheon}, \citenamefont {Lassen},
  \citenamefont {Martin}, \citenamefont {Ottewell}, \citenamefont {Ouellet},
  \citenamefont {Ruprecht}, \citenamefont {Setoodehnia}, \citenamefont
  {Shotter}, \citenamefont {Teigelh\"ofer}, \citenamefont {Vockenhuber},
  \citenamefont {Wrede},\ and\ \citenamefont {Wallner}}]{Erikson}%
  \BibitemOpen
  \bibfield  {author} {\bibinfo {author} {\bibfnamefont {L.}~\bibnamefont
  {Erikson}}, \bibinfo {author} {\bibfnamefont {C.}~\bibnamefont {Ruiz}},
  \bibinfo {author} {\bibfnamefont {F.}~\bibnamefont {Ames}}, \bibinfo {author}
  {\bibfnamefont {P.}~\bibnamefont {Bricault}}, \bibinfo {author}
  {\bibfnamefont {L.}~\bibnamefont {Buchmann}}, \bibinfo {author}
  {\bibfnamefont {A.~A.}\ \bibnamefont {Chen}}, \bibinfo {author}
  {\bibfnamefont {J.}~\bibnamefont {Chen}}, \bibinfo {author} {\bibfnamefont
  {H.}~\bibnamefont {Dare}}, \bibinfo {author} {\bibfnamefont {B.}~\bibnamefont
  {Davids}}, \bibinfo {author} {\bibfnamefont {C.}~\bibnamefont {Davis}},
  \bibinfo {author} {\bibfnamefont {C.~M.}\ \bibnamefont {Deibel}}, \bibinfo
  {author} {\bibfnamefont {M.}~\bibnamefont {Dombsky}}, \bibinfo {author}
  {\bibfnamefont {S.}~\bibnamefont {Foubister}}, \bibinfo {author}
  {\bibfnamefont {N.}~\bibnamefont {Galinski}}, \bibinfo {author}
  {\bibfnamefont {U.}~\bibnamefont {Greife}}, \bibinfo {author} {\bibfnamefont
  {U.}~\bibnamefont {Hager}}, \bibinfo {author} {\bibfnamefont
  {A.}~\bibnamefont {Hussein}}, \bibinfo {author} {\bibfnamefont {D.~A.}\
  \bibnamefont {Hutcheon}}, \bibinfo {author} {\bibfnamefont {J.}~\bibnamefont
  {Lassen}}, \bibinfo {author} {\bibfnamefont {L.}~\bibnamefont {Martin}},
  \bibinfo {author} {\bibfnamefont {D.~F.}\ \bibnamefont {Ottewell}}, \bibinfo
  {author} {\bibfnamefont {C.~V.}\ \bibnamefont {Ouellet}}, \bibinfo {author}
  {\bibfnamefont {G.}~\bibnamefont {Ruprecht}}, \bibinfo {author}
  {\bibfnamefont {K.}~\bibnamefont {Setoodehnia}}, \bibinfo {author}
  {\bibfnamefont {A.~C.}\ \bibnamefont {Shotter}}, \bibinfo {author}
  {\bibfnamefont {A.}~\bibnamefont {Teigelh\"ofer}}, \bibinfo {author}
  {\bibfnamefont {C.}~\bibnamefont {Vockenhuber}}, \bibinfo {author}
  {\bibfnamefont {C.}~\bibnamefont {Wrede}}, \ and\ \bibinfo {author}
  {\bibfnamefont {A.}~\bibnamefont {Wallner}},\ }\href {\doibase
  10.1103/PhysRevC.81.045808} {\bibfield  {journal} {\bibinfo  {journal} {Phys.
  Rev. C}\ }\textbf {\bibinfo {volume} {81}},\ \bibinfo {pages} {045808}
  (\bibinfo {year} {2010})}\BibitemShut {NoStop}%
\bibitem [{\citenamefont {Hutcheon}\ \emph {et~al.}(2003)\citenamefont
  {Hutcheon}, \citenamefont {Bishop}, \citenamefont {Buchmann}, \citenamefont
  {Chatterjee}, \citenamefont {Chen}, \citenamefont {D'Auria}, \citenamefont
  {Engel}, \citenamefont {Gigliotti}, \citenamefont {Greife}, \citenamefont
  {Hunter}, \citenamefont {Hussein}, \citenamefont {Jewett}, \citenamefont
  {Khan}, \citenamefont {Lamey}, \citenamefont {Laird}, \citenamefont {Liu},
  \citenamefont {Olin}, \citenamefont {Ottewell}, \citenamefont {Rogers},
  \citenamefont {Roy}, \citenamefont {Sprenger},\ and\ \citenamefont
  {Wrede}}]{dragon}%
  \BibitemOpen
  \bibfield  {author} {\bibinfo {author} {\bibfnamefont {D.}~\bibnamefont
  {Hutcheon}}, \bibinfo {author} {\bibfnamefont {S.}~\bibnamefont {Bishop}},
  \bibinfo {author} {\bibfnamefont {L.}~\bibnamefont {Buchmann}}, \bibinfo
  {author} {\bibfnamefont {M.}~\bibnamefont {Chatterjee}}, \bibinfo {author}
  {\bibfnamefont {A.}~\bibnamefont {Chen}}, \bibinfo {author} {\bibfnamefont
  {J.}~\bibnamefont {D'Auria}}, \bibinfo {author} {\bibfnamefont
  {S.}~\bibnamefont {Engel}}, \bibinfo {author} {\bibfnamefont
  {D.}~\bibnamefont {Gigliotti}}, \bibinfo {author} {\bibfnamefont
  {U.}~\bibnamefont {Greife}}, \bibinfo {author} {\bibfnamefont
  {D.}~\bibnamefont {Hunter}}, \bibinfo {author} {\bibfnamefont
  {A.}~\bibnamefont {Hussein}}, \bibinfo {author} {\bibfnamefont
  {C.}~\bibnamefont {Jewett}}, \bibinfo {author} {\bibfnamefont
  {N.}~\bibnamefont {Khan}}, \bibinfo {author} {\bibfnamefont {M.}~\bibnamefont
  {Lamey}}, \bibinfo {author} {\bibfnamefont {A.}~\bibnamefont {Laird}},
  \bibinfo {author} {\bibfnamefont {W.}~\bibnamefont {Liu}}, \bibinfo {author}
  {\bibfnamefont {A.}~\bibnamefont {Olin}}, \bibinfo {author} {\bibfnamefont
  {D.}~\bibnamefont {Ottewell}}, \bibinfo {author} {\bibfnamefont
  {J.}~\bibnamefont {Rogers}}, \bibinfo {author} {\bibfnamefont
  {G.}~\bibnamefont {Roy}}, \bibinfo {author} {\bibfnamefont {H.}~\bibnamefont
  {Sprenger}}, \ and\ \bibinfo {author} {\bibfnamefont {C.}~\bibnamefont
  {Wrede}},\ }\href {\doibase https://doi.org/10.1016/S0168-9002(02)01990-3}
  {\bibfield  {journal} {\bibinfo  {journal} {Nucl. Instr. Meth. Phys. Res. A}\
  }\textbf {\bibinfo {volume} {498}},\ \bibinfo {pages} {190} (\bibinfo {year}
  {2003})}\BibitemShut {NoStop}%
\bibitem [{\citenamefont {Lotay}\ \emph {et~al.}(2008)\citenamefont {Lotay},
  \citenamefont {Woods}, \citenamefont {Seweryniak}, \citenamefont {Carpenter},
  \citenamefont {Hoteling}, \citenamefont {Janssens}, \citenamefont {Jenkins},
  \citenamefont {Lauritsen}, \citenamefont {Lister}, \citenamefont {Robinson},\
  and\ \citenamefont {Zhu}}]{G_Lotay}%
  \BibitemOpen
  \bibfield  {author} {\bibinfo {author} {\bibfnamefont {G.}~\bibnamefont
  {Lotay}}, \bibinfo {author} {\bibfnamefont {P.~J.}\ \bibnamefont {Woods}},
  \bibinfo {author} {\bibfnamefont {D.}~\bibnamefont {Seweryniak}}, \bibinfo
  {author} {\bibfnamefont {M.~P.}\ \bibnamefont {Carpenter}}, \bibinfo {author}
  {\bibfnamefont {N.}~\bibnamefont {Hoteling}}, \bibinfo {author}
  {\bibfnamefont {R.~V.~F.}\ \bibnamefont {Janssens}}, \bibinfo {author}
  {\bibfnamefont {D.~G.}\ \bibnamefont {Jenkins}}, \bibinfo {author}
  {\bibfnamefont {T.}~\bibnamefont {Lauritsen}}, \bibinfo {author}
  {\bibfnamefont {C.~J.}\ \bibnamefont {Lister}}, \bibinfo {author}
  {\bibfnamefont {A.}~\bibnamefont {Robinson}}, \ and\ \bibinfo {author}
  {\bibfnamefont {S.}~\bibnamefont {Zhu}},\ }\href {\doibase
  10.1103/PhysRevC.77.042802} {\bibfield  {journal} {\bibinfo  {journal} {Phys.
  Rev. C}\ }\textbf {\bibinfo {volume} {77}},\ \bibinfo {pages} {042802}
  (\bibinfo {year} {2008})}\BibitemShut {NoStop}%
\bibitem [{\citenamefont {Neveling}\ \emph {et~al.}(2011)\citenamefont
  {Neveling}, \citenamefont {Fujita}, \citenamefont {Smit}, \citenamefont
  {Adachi}, \citenamefont {Berg}, \citenamefont {Buthelezi}, \citenamefont
  {Carter}, \citenamefont {Conradie}, \citenamefont {Couder}, \citenamefont
  {Fearick} \emph {et~al.}}]{k600}%
  \BibitemOpen
  \bibfield  {author} {\bibinfo {author} {\bibfnamefont {R.}~\bibnamefont
  {Neveling}}, \bibinfo {author} {\bibfnamefont {H.}~\bibnamefont {Fujita}},
  \bibinfo {author} {\bibfnamefont {F.}~\bibnamefont {Smit}}, \bibinfo {author}
  {\bibfnamefont {T.}~\bibnamefont {Adachi}}, \bibinfo {author} {\bibfnamefont
  {G.}~\bibnamefont {Berg}}, \bibinfo {author} {\bibfnamefont {E.}~\bibnamefont
  {Buthelezi}}, \bibinfo {author} {\bibfnamefont {J.}~\bibnamefont {Carter}},
  \bibinfo {author} {\bibfnamefont {J.}~\bibnamefont {Conradie}}, \bibinfo
  {author} {\bibfnamefont {M.}~\bibnamefont {Couder}}, \bibinfo {author}
  {\bibfnamefont {R.}~\bibnamefont {Fearick}},  \emph {et~al.},\ }\href
  {\doibase 10.1016/j.nima.2011.06.077} {\bibfield  {journal} {\bibinfo
  {journal} {Nucl. Instr. Meth. Phys. Res. A}\ }\textbf {\bibinfo {volume}
  {654}},\ \bibinfo {pages} {29} (\bibinfo {year} {2011})}\BibitemShut
  {NoStop}%
\bibitem [{\citenamefont {Adsley}\ \emph {et~al.}(2017)\citenamefont {Adsley},
  \citenamefont {Neveling}, \citenamefont {Papka}, \citenamefont {Dyers},
  \citenamefont {BrÃ¼mmer}, \citenamefont {Diget}, \citenamefont {Hubbard},
  \citenamefont {Li}, \citenamefont {Long}, \citenamefont {Marin-Lambarri},
  \citenamefont {Pellegri}, \citenamefont {Pesudo}, \citenamefont {Pool},
  \citenamefont {Smit},\ and\ \citenamefont {Triambak}}]{Adsley_2017}%
  \BibitemOpen
  \bibfield  {author} {\bibinfo {author} {\bibfnamefont {P.}~\bibnamefont
  {Adsley}}, \bibinfo {author} {\bibfnamefont {R.}~\bibnamefont {Neveling}},
  \bibinfo {author} {\bibfnamefont {P.}~\bibnamefont {Papka}}, \bibinfo
  {author} {\bibfnamefont {Z.}~\bibnamefont {Dyers}}, \bibinfo {author}
  {\bibfnamefont {J.}~\bibnamefont {BrÃ¼mmer}}, \bibinfo {author}
  {\bibfnamefont {C.}~\bibnamefont {Diget}}, \bibinfo {author} {\bibfnamefont
  {N.}~\bibnamefont {Hubbard}}, \bibinfo {author} {\bibfnamefont
  {K.}~\bibnamefont {Li}}, \bibinfo {author} {\bibfnamefont {A.}~\bibnamefont
  {Long}}, \bibinfo {author} {\bibfnamefont {D.}~\bibnamefont
  {Marin-Lambarri}}, \bibinfo {author} {\bibfnamefont {L.}~\bibnamefont
  {Pellegri}}, \bibinfo {author} {\bibfnamefont {V.}~\bibnamefont {Pesudo}},
  \bibinfo {author} {\bibfnamefont {L.}~\bibnamefont {Pool}}, \bibinfo {author}
  {\bibfnamefont {F.}~\bibnamefont {Smit}}, \ and\ \bibinfo {author}
  {\bibfnamefont {S.}~\bibnamefont {Triambak}},\ }\href {\doibase
  10.1088/1748-0221/12/02/t02004} {\bibfield  {journal} {\bibinfo  {journal}
  {JINST}\ }\textbf {\bibinfo {volume} {12}},\ \bibinfo {pages} {T02004}
  (\bibinfo {year} {2017})}\BibitemShut {NoStop}%
\bibitem [{\citenamefont {Triambak}\ \emph {et~al.}(2006)\citenamefont
  {Triambak}, \citenamefont {Garc\'{\i}a}, \citenamefont {Melconian},
  \citenamefont {Mella},\ and\ \citenamefont {Biesel}}]{Triambak}%
  \BibitemOpen
  \bibfield  {author} {\bibinfo {author} {\bibfnamefont {S.}~\bibnamefont
  {Triambak}}, \bibinfo {author} {\bibfnamefont {A.}~\bibnamefont
  {Garc\'{\i}a}}, \bibinfo {author} {\bibfnamefont {D.}~\bibnamefont
  {Melconian}}, \bibinfo {author} {\bibfnamefont {M.}~\bibnamefont {Mella}}, \
  and\ \bibinfo {author} {\bibfnamefont {O.}~\bibnamefont {Biesel}},\ }\href
  {\doibase 10.1103/PhysRevC.74.054306} {\bibfield  {journal} {\bibinfo
  {journal} {Phys. Rev. C}\ }\textbf {\bibinfo {volume} {74}},\ \bibinfo
  {pages} {054306} (\bibinfo {year} {2006})}\BibitemShut {NoStop}%
\bibitem [{\citenamefont {Kamil}\ \emph {et~al.}(2022)\citenamefont {Kamil},
  \citenamefont {Triambak}, \citenamefont {Ball}, \citenamefont {Bildstein},
  \citenamefont {Varela}, \citenamefont {Faestermann}, \citenamefont {Garrett},
  \citenamefont {Moradi}, \citenamefont {Hertenberger}, \citenamefont {Kheswa},
  \citenamefont {Mukwevho}, \citenamefont {Rebeiro},\ and\ \citenamefont
  {Wirth}}]{Kamil}%
  \BibitemOpen
  \bibfield  {author} {\bibinfo {author} {\bibfnamefont {M.}~\bibnamefont
  {Kamil}}, \bibinfo {author} {\bibfnamefont {S.}~\bibnamefont {Triambak}},
  \bibinfo {author} {\bibfnamefont {G.~C.}\ \bibnamefont {Ball}}, \bibinfo
  {author} {\bibfnamefont {V.}~\bibnamefont {Bildstein}}, \bibinfo {author}
  {\bibfnamefont {A.~D.}\ \bibnamefont {Varela}}, \bibinfo {author}
  {\bibfnamefont {T.}~\bibnamefont {Faestermann}}, \bibinfo {author}
  {\bibfnamefont {P.~E.}\ \bibnamefont {Garrett}}, \bibinfo {author}
  {\bibfnamefont {F.~G.}\ \bibnamefont {Moradi}}, \bibinfo {author}
  {\bibfnamefont {R.}~\bibnamefont {Hertenberger}}, \bibinfo {author}
  {\bibfnamefont {N.~Y.}\ \bibnamefont {Kheswa}}, \bibinfo {author}
  {\bibfnamefont {N.~J.}\ \bibnamefont {Mukwevho}}, \bibinfo {author}
  {\bibfnamefont {B.~M.}\ \bibnamefont {Rebeiro}}, \ and\ \bibinfo {author}
  {\bibfnamefont {H.-F.}\ \bibnamefont {Wirth}},\ }\href {\doibase
  10.1103/PhysRevC.105.055805} {\bibfield  {journal} {\bibinfo  {journal}
  {Phys. Rev. C}\ }\textbf {\bibinfo {volume} {105}},\ \bibinfo {pages}
  {055805} (\bibinfo {year} {2022})}\BibitemShut {NoStop}%
\bibitem [{\citenamefont {Honkanen}\ \emph {et~al.}(1979)\citenamefont
  {Honkanen}, \citenamefont {Kortelahti}, \citenamefont {\"Ayst\"o},
  \citenamefont {Eskola},\ and\ \citenamefont {Hautoj\"arvi}}]{Honkanen}%
  \BibitemOpen
  \bibfield  {author} {\bibinfo {author} {\bibfnamefont {J.}~\bibnamefont
  {Honkanen}}, \bibinfo {author} {\bibfnamefont {M.}~\bibnamefont
  {Kortelahti}}, \bibinfo {author} {\bibfnamefont {J.}~\bibnamefont
  {\"Ayst\"o}}, \bibinfo {author} {\bibfnamefont {K.}~\bibnamefont {Eskola}}, \
  and\ \bibinfo {author} {\bibfnamefont {A.}~\bibnamefont {Hautoj\"arvi}},\
  }\href {\doibase 10.1088/0031-8949/19/3/004} {\bibfield  {journal} {\bibinfo
  {journal} {Physica Scripta}\ }\textbf {\bibinfo {volume} {19}},\ \bibinfo
  {pages} {239} (\bibinfo {year} {1979})}\BibitemShut {NoStop}%
\bibitem [{\citenamefont {Wang}\ \emph {et~al.}(2021)\citenamefont {Wang},
  \citenamefont {Huang}, \citenamefont {Kondev}, \citenamefont {Audi},\ and\
  \citenamefont {Naimi}}]{ame2021}%
  \BibitemOpen
  \bibfield  {author} {\bibinfo {author} {\bibfnamefont {M.}~\bibnamefont
  {Wang}}, \bibinfo {author} {\bibfnamefont {W.}~\bibnamefont {Huang}},
  \bibinfo {author} {\bibfnamefont {F.}~\bibnamefont {Kondev}}, \bibinfo
  {author} {\bibfnamefont {G.}~\bibnamefont {Audi}}, \ and\ \bibinfo {author}
  {\bibfnamefont {S.}~\bibnamefont {Naimi}},\ }\href@noop {} {\bibfield
  {journal} {\bibinfo  {journal} {Chin. Phys. C}\ }\textbf {\bibinfo {volume}
  {45}},\ \bibinfo {pages} {030003} (\bibinfo {year} {2021})}\BibitemShut
  {NoStop}%
\bibitem [{\citenamefont {Audi}\ \emph {et~al.}(2003)\citenamefont {Audi},
  \citenamefont {Wapstra},\ and\ \citenamefont {Thibault}}]{masses:03}%
  \BibitemOpen
  \bibfield  {author} {\bibinfo {author} {\bibfnamefont {G.}~\bibnamefont
  {Audi}}, \bibinfo {author} {\bibfnamefont {A.}~\bibnamefont {Wapstra}}, \
  and\ \bibinfo {author} {\bibfnamefont {C.}~\bibnamefont {Thibault}},\ }\href
  {\doibase https://doi.org/10.1016/j.nuclphysa.2003.11.003} {\bibfield
  {journal} {\bibinfo  {journal} {Nuclear Physics A}\ }\textbf {\bibinfo
  {volume} {729}},\ \bibinfo {pages} {337} (\bibinfo {year} {2003})},\ \bibinfo
  {note} {the 2003 NUBASE and Atomic Mass Evaluations}\BibitemShut {NoStop}%
\bibitem [{\citenamefont {Zegers}\ \emph {et~al.}(2008)\citenamefont {Zegers},
  \citenamefont {Meharchand}, \citenamefont {Adachi}, \citenamefont {Austin},
  \citenamefont {Brown}, \citenamefont {Fujita}, \citenamefont {Fujiwara},
  \citenamefont {Guess}, \citenamefont {Hashimoto}, \citenamefont {Hatanaka},
  \citenamefont {Howard}, \citenamefont {Matsubara}, \citenamefont {Nakanishi},
  \citenamefont {Ohta}, \citenamefont {Okamura}, \citenamefont {Sakemi},
  \citenamefont {Shimbara}, \citenamefont {Shimizu}, \citenamefont {Scholl},
  \citenamefont {Signoracci}, \citenamefont {Tameshige}, \citenamefont
  {Tamii},\ and\ \citenamefont {Yosoi}}]{Zegers}%
  \BibitemOpen
  \bibfield  {author} {\bibinfo {author} {\bibfnamefont {R.~G.~T.}\
  \bibnamefont {Zegers}}, \bibinfo {author} {\bibfnamefont {R.}~\bibnamefont
  {Meharchand}}, \bibinfo {author} {\bibfnamefont {T.}~\bibnamefont {Adachi}},
  \bibinfo {author} {\bibfnamefont {S.~M.}\ \bibnamefont {Austin}}, \bibinfo
  {author} {\bibfnamefont {B.~A.}\ \bibnamefont {Brown}}, \bibinfo {author}
  {\bibfnamefont {Y.}~\bibnamefont {Fujita}}, \bibinfo {author} {\bibfnamefont
  {M.}~\bibnamefont {Fujiwara}}, \bibinfo {author} {\bibfnamefont {C.~J.}\
  \bibnamefont {Guess}}, \bibinfo {author} {\bibfnamefont {H.}~\bibnamefont
  {Hashimoto}}, \bibinfo {author} {\bibfnamefont {K.}~\bibnamefont {Hatanaka}},
  \bibinfo {author} {\bibfnamefont {M.~E.}\ \bibnamefont {Howard}}, \bibinfo
  {author} {\bibfnamefont {H.}~\bibnamefont {Matsubara}}, \bibinfo {author}
  {\bibfnamefont {K.}~\bibnamefont {Nakanishi}}, \bibinfo {author}
  {\bibfnamefont {T.}~\bibnamefont {Ohta}}, \bibinfo {author} {\bibfnamefont
  {H.}~\bibnamefont {Okamura}}, \bibinfo {author} {\bibfnamefont
  {Y.}~\bibnamefont {Sakemi}}, \bibinfo {author} {\bibfnamefont
  {Y.}~\bibnamefont {Shimbara}}, \bibinfo {author} {\bibfnamefont
  {Y.}~\bibnamefont {Shimizu}}, \bibinfo {author} {\bibfnamefont
  {C.}~\bibnamefont {Scholl}}, \bibinfo {author} {\bibfnamefont
  {A.}~\bibnamefont {Signoracci}}, \bibinfo {author} {\bibfnamefont
  {Y.}~\bibnamefont {Tameshige}}, \bibinfo {author} {\bibfnamefont
  {A.}~\bibnamefont {Tamii}}, \ and\ \bibinfo {author} {\bibfnamefont
  {M.}~\bibnamefont {Yosoi}},\ }\href {\doibase 10.1103/PhysRevC.78.014314}
  {\bibfield  {journal} {\bibinfo  {journal} {Phys. Rev. C}\ }\textbf {\bibinfo
  {volume} {78}},\ \bibinfo {pages} {014314} (\bibinfo {year}
  {2008})}\BibitemShut {NoStop}%
\bibitem [{\citenamefont {Visser}\ \emph {et~al.}(2008)\citenamefont {Visser},
  \citenamefont {Wrede}, \citenamefont {Caggiano}, \citenamefont {Clark},
  \citenamefont {Deibel}, \citenamefont {Lewis}, \citenamefont {Parikh},\ and\
  \citenamefont {Parker}}]{Visser2}%
  \BibitemOpen
  \bibfield  {author} {\bibinfo {author} {\bibfnamefont {D.~W.}\ \bibnamefont
  {Visser}}, \bibinfo {author} {\bibfnamefont {C.}~\bibnamefont {Wrede}},
  \bibinfo {author} {\bibfnamefont {J.~A.}\ \bibnamefont {Caggiano}}, \bibinfo
  {author} {\bibfnamefont {J.~A.}\ \bibnamefont {Clark}}, \bibinfo {author}
  {\bibfnamefont {C.~M.}\ \bibnamefont {Deibel}}, \bibinfo {author}
  {\bibfnamefont {R.}~\bibnamefont {Lewis}}, \bibinfo {author} {\bibfnamefont
  {A.}~\bibnamefont {Parikh}}, \ and\ \bibinfo {author} {\bibfnamefont {P.~D.}\
  \bibnamefont {Parker}},\ }\href {\doibase 10.1103/PhysRevC.78.028802}
  {\bibfield  {journal} {\bibinfo  {journal} {Phys. Rev. C}\ }\textbf {\bibinfo
  {volume} {78}},\ \bibinfo {pages} {028802} (\bibinfo {year}
  {2008})}\BibitemShut {NoStop}%
\bibitem [{\citenamefont {No\'e}\ \emph {et~al.}(1974)\citenamefont {No\'e},
  \citenamefont {Balamuth},\ and\ \citenamefont {Zurm\"uhle}}]{Noe:1974}%
  \BibitemOpen
  \bibfield  {author} {\bibinfo {author} {\bibfnamefont {J.~W.}\ \bibnamefont
  {No\'e}}, \bibinfo {author} {\bibfnamefont {D.~P.}\ \bibnamefont {Balamuth}},
  \ and\ \bibinfo {author} {\bibfnamefont {R.~W.}\ \bibnamefont {Zurm\"uhle}},\
  }\href {\doibase 10.1103/PhysRevC.9.132} {\bibfield  {journal} {\bibinfo
  {journal} {Phys. Rev. C}\ }\textbf {\bibinfo {volume} {9}},\ \bibinfo {pages}
  {132} (\bibinfo {year} {1974})}\BibitemShut {NoStop}%
\bibitem [{NND()}]{NNDC}%
  \BibitemOpen
  \href@noop {} {}\bibinfo {howpublished}
  {\url{https://www.nndc.bnl.gov}}\BibitemShut {NoStop}%
\bibitem [{\citenamefont {Lee}\ \emph {et~al.}(2007)\citenamefont {Lee},
  \citenamefont {Per\"aj\"arvi}, \citenamefont {Powell}, \citenamefont
  {O'Neil}, \citenamefont {Moltz}, \citenamefont {Goldberg},\ and\
  \citenamefont {Cerny}}]{Lee:2007}%
  \BibitemOpen
  \bibfield  {author} {\bibinfo {author} {\bibfnamefont {D.~W.}\ \bibnamefont
  {Lee}}, \bibinfo {author} {\bibfnamefont {K.}~\bibnamefont {Per\"aj\"arvi}},
  \bibinfo {author} {\bibfnamefont {J.}~\bibnamefont {Powell}}, \bibinfo
  {author} {\bibfnamefont {J.~P.}\ \bibnamefont {O'Neil}}, \bibinfo {author}
  {\bibfnamefont {D.~M.}\ \bibnamefont {Moltz}}, \bibinfo {author}
  {\bibfnamefont {V.~Z.}\ \bibnamefont {Goldberg}}, \ and\ \bibinfo {author}
  {\bibfnamefont {J.}~\bibnamefont {Cerny}},\ }\href {\doibase
  10.1103/PhysRevC.76.024314} {\bibfield  {journal} {\bibinfo  {journal} {Phys.
  Rev. C}\ }\textbf {\bibinfo {volume} {76}},\ \bibinfo {pages} {024314}
  (\bibinfo {year} {2007})}\BibitemShut {NoStop}%
\bibitem [{\citenamefont {Ormand}\ and\ \citenamefont
  {Brown}(1989)}]{Ormand_Brown}%
  \BibitemOpen
  \bibfield  {author} {\bibinfo {author} {\bibfnamefont {W.~E.}\ \bibnamefont
  {Ormand}}\ and\ \bibinfo {author} {\bibfnamefont {B.~A.}\ \bibnamefont
  {Brown}},\ }\href {\doibase https://doi.org/10.1016/0375-9474(89)90203-0}
  {\bibfield  {journal} {\bibinfo  {journal} {Nucl. Phys. A}\ }\textbf
  {\bibinfo {volume} {491}},\ \bibinfo {pages} {1} (\bibinfo {year}
  {1989})}\BibitemShut {NoStop}%
\bibitem [{\citenamefont {Magilligan}\ and\ \citenamefont
  {Brown}(2020)}]{usdc}%
  \BibitemOpen
  \bibfield  {author} {\bibinfo {author} {\bibfnamefont {A.}~\bibnamefont
  {Magilligan}}\ and\ \bibinfo {author} {\bibfnamefont {B.~A.}\ \bibnamefont
  {Brown}},\ }\href {\doibase 10.1103/PhysRevC.101.064312} {\bibfield
  {journal} {\bibinfo  {journal} {Phys. Rev. C}\ }\textbf {\bibinfo {volume}
  {101}},\ \bibinfo {pages} {064312} (\bibinfo {year} {2020})}\BibitemShut
  {NoStop}%
\bibitem [{\citenamefont {Richter}\ \emph {et~al.}(2008)\citenamefont
  {Richter}, \citenamefont {Mkhize},\ and\ \citenamefont {Brown}}]{Richter:08}%
  \BibitemOpen
  \bibfield  {author} {\bibinfo {author} {\bibfnamefont {W.~A.}\ \bibnamefont
  {Richter}}, \bibinfo {author} {\bibfnamefont {S.}~\bibnamefont {Mkhize}}, \
  and\ \bibinfo {author} {\bibfnamefont {B.~A.}\ \bibnamefont {Brown}},\ }\href
  {\doibase 10.1103/PhysRevC.78.064302} {\bibfield  {journal} {\bibinfo
  {journal} {Phys. Rev. C}\ }\textbf {\bibinfo {volume} {78}},\ \bibinfo
  {pages} {064302} (\bibinfo {year} {2008})}\BibitemShut {NoStop}%
\bibitem [{\citenamefont {Wrede}\ \emph {et~al.}(2010)\citenamefont {Wrede},
  \citenamefont {Clark}, \citenamefont {Deibel}, \citenamefont {Faestermann},
  \citenamefont {Hertenberger}, \citenamefont {Parikh}, \citenamefont {Wirth},
  \citenamefont {Bishop}, \citenamefont {Chen}, \citenamefont {Eppinger},
  \citenamefont {Freeman}, \citenamefont {Kr\"ucken}, \citenamefont
  {Lepyoshkina}, \citenamefont {Rugel},\ and\ \citenamefont
  {Setoodehnia}}]{Wrede}%
  \BibitemOpen
  \bibfield  {author} {\bibinfo {author} {\bibfnamefont {C.}~\bibnamefont
  {Wrede}}, \bibinfo {author} {\bibfnamefont {J.~A.}\ \bibnamefont {Clark}},
  \bibinfo {author} {\bibfnamefont {C.~M.}\ \bibnamefont {Deibel}}, \bibinfo
  {author} {\bibfnamefont {T.}~\bibnamefont {Faestermann}}, \bibinfo {author}
  {\bibfnamefont {R.}~\bibnamefont {Hertenberger}}, \bibinfo {author}
  {\bibfnamefont {A.}~\bibnamefont {Parikh}}, \bibinfo {author} {\bibfnamefont
  {H.-F.}\ \bibnamefont {Wirth}}, \bibinfo {author} {\bibfnamefont
  {S.}~\bibnamefont {Bishop}}, \bibinfo {author} {\bibfnamefont {A.~A.}\
  \bibnamefont {Chen}}, \bibinfo {author} {\bibfnamefont {K.}~\bibnamefont
  {Eppinger}}, \bibinfo {author} {\bibfnamefont {B.~M.}\ \bibnamefont
  {Freeman}}, \bibinfo {author} {\bibfnamefont {R.}~\bibnamefont {Kr\"ucken}},
  \bibinfo {author} {\bibfnamefont {O.}~\bibnamefont {Lepyoshkina}}, \bibinfo
  {author} {\bibfnamefont {G.}~\bibnamefont {Rugel}}, \ and\ \bibinfo {author}
  {\bibfnamefont {K.}~\bibnamefont {Setoodehnia}},\ }\href {\doibase
  10.1103/PhysRevC.82.035805} {\bibfield  {journal} {\bibinfo  {journal} {Phys.
  Rev. C}\ }\textbf {\bibinfo {volume} {82}},\ \bibinfo {pages} {035805}
  (\bibinfo {year} {2010})}\BibitemShut {NoStop}%
\bibitem [{\citenamefont {Longland}\ \emph {et~al.}(2010)\citenamefont
  {Longland}, \citenamefont {Iliadis}, \citenamefont {Champagne}, \citenamefont
  {Newton}, \citenamefont {Ugalde}, \citenamefont {Coc},\ and\ \citenamefont
  {Fitzgerald}}]{RatesMC1}%
  \BibitemOpen
  \bibfield  {author} {\bibinfo {author} {\bibfnamefont {R.}~\bibnamefont
  {Longland}}, \bibinfo {author} {\bibfnamefont {C.}~\bibnamefont {Iliadis}},
  \bibinfo {author} {\bibfnamefont {A.}~\bibnamefont {Champagne}}, \bibinfo
  {author} {\bibfnamefont {J.~R.}\ \bibnamefont {Newton}}, \bibinfo {author}
  {\bibfnamefont {C.}~\bibnamefont {Ugalde}}, \bibinfo {author} {\bibfnamefont
  {A.}~\bibnamefont {Coc}}, \ and\ \bibinfo {author} {\bibfnamefont
  {R.}~\bibnamefont {Fitzgerald}},\ }\href@noop {} {\bibfield  {journal}
  {\bibinfo  {journal} {Nucl. Phys. A}\ }\textbf {\bibinfo {volume} {841}},\
  \bibinfo {pages} {1} (\bibinfo {year} {2010})}\BibitemShut {NoStop}%
\bibitem [{\citenamefont {Iliadis}\ \emph {et~al.}(2010)\citenamefont
  {Iliadis}, \citenamefont {Longland}, \citenamefont {Champagne}, \citenamefont
  {Coc},\ and\ \citenamefont {Fitzgerald}}]{RatesMC2}%
  \BibitemOpen
  \bibfield  {author} {\bibinfo {author} {\bibfnamefont {C.}~\bibnamefont
  {Iliadis}}, \bibinfo {author} {\bibfnamefont {R.}~\bibnamefont {Longland}},
  \bibinfo {author} {\bibfnamefont {A.}~\bibnamefont {Champagne}}, \bibinfo
  {author} {\bibfnamefont {A.}~\bibnamefont {Coc}}, \ and\ \bibinfo {author}
  {\bibfnamefont {R.}~\bibnamefont {Fitzgerald}},\ }\href@noop {} {\bibfield
  {journal} {\bibinfo  {journal} {Nucl. Phys. A}\ }\textbf {\bibinfo {volume}
  {841}},\ \bibinfo {pages} {31} (\bibinfo {year} {2010})}\BibitemShut
  {NoStop}%
\end{thebibliography}%

\end{document}